\documentclass[letterpaper,peerreview,draftclsnofoot,onecolumn,11pt]{IEEEtran}
\usepackage{subfig, amsfonts,amsmath, amssymb, amsthm, graphicx, cite}
\usepackage{bm}
\usepackage{algorithm,algorithmic}
\usepackage[latin1]{inputenc}
\usepackage{amsmath}
\usepackage{amssymb}
\usepackage{graphics}
\usepackage{graphicx}
\usepackage{epstopdf}
\usepackage{latexsym}
\usepackage{amsfonts}
\usepackage{color}
\usepackage[table]{xcolor}
\usepackage{url}
\usepackage{amsmath}
\usepackage{multirow}
\usepackage{tikz}
\usetikzlibrary{positioning,chains,fit,shapes,calc} 
\usetikzlibrary{shapes,arrows}
\usetikzlibrary{arrows}
\usepackage{color}
\usepackage{scalefnt}
\usepackage{textcomp}
\usepackage{xcolor}
\usepackage{booktabs}

\newtheorem{theorem}{Theorem}

\newtheorem{lemma}[theorem]{Lemma}
\newtheorem{proposition}[theorem]{Proposition}

\newtheorem*{case*}{Case}
\theoremstyle{definition}
\newtheorem{definition}{Definition}

\definecolor{myblue}{RGB}{0,0,100}
\definecolor{mygreen}{RGB}{0,0,100}
\definecolor{myyellow}{RGB}{0,100,100}
\definecolor{my1}{RGB}{100,0,100}
\definecolor{myred}{RGB}{100,0,0}
\definecolor{my2}{RGB}{0,100,0}

\begin{document}

\title{Universal and Succinct Source Coding of Deep Neural Networks}

\author{Sourya~Basu
        and~Lav~R.~Varshney,~\IEEEmembership{Senior Member,~IEEE}
\thanks{S.~Basu was with the Department of Electrical Engineering, Indian Institute of Technology Kanpur, Kanpur 208016, India and is with the Coordinated Science Laboratory and the Department of Electrical and Computer Engineering, University of Illinois at Urbana-Champaign, Urbana, IL 61801, USA (e-mail: sourya@illinois.edu).}%
\thanks{L.~R. Varshney is with the Coordinated Science Laboratory and the Department of Electrical and Computer Engineering, University of Illinois at Urbana-Champaign, Urbana, IL 61801, USA (e-mail: varshney@illinois.edu).}
\thanks{The material in this paper was presented in part at the 2017 IEEE Data Compression Conference \cite{BasuV2017}.}
\thanks{This work was supported in part by Systems on Nanoscale Information fabriCs (SONIC), one of the six SRC STARnet Centers, sponsored by MARCO and DARPA, and in part by grant number 2018-182794 from the Chan Zuckerberg Initiative DAF, an advised fund of Silicon Valley Community Foundation.}
}

\maketitle

\begin{abstract}
Deep neural networks have shown incredible performance for inference tasks in a variety of domains.  Unfortunately, most current deep networks are enormous cloud-based structures that require significant storage space, which limits scaling of deep learning as a service (DLaaS) and use for on-device intelligence.  This paper is concerned with finding universal lossless compressed representations of deep feedforward networks with synaptic weights drawn from discrete sets, and directly performing inference without full decompression.  The basic insight that allows less rate than na\"{i}ve approaches is recognizing that the bipartite graph layers of feedforward networks have a kind of permutation invariance to the labeling of nodes, in terms of inferential operation.
We provide efficient algorithms to dissipate this irrelevant uncertainty and then use arithmetic coding to nearly achieve the entropy bound in a universal manner. We also provide experimental results of our approach on several standard datasets.
\end{abstract}


\section{Introduction}
\label{sec:intro}

Deep learning has achieved incredible performance for inference tasks such as speech recognition, image recognition, and natural language processing.  Most current deep neural networks, however, are enormous cloud-based structures that are \emph{too large} and \emph{too complex} to perform fast, energy-efficient inference on device. Even in the cloud, providing personalized deep learning as a service (DLaaS), where each customer for an application like bank fraud detection may require a different trained network, scaling to millions of stored networks is not possible.  Compression, with the capability of providing inference without full decompression, is important. We develop new universal source coding techniques for feedforward deep networks having synaptic weights drawn from finite sets that essentially achieve the entropy lower bound, which we also compute. Further, we provide an algorithm to use these compressed representations for inference tasks without complete decompression.
Structures that can represent information near the entropy bound while also allowing efficient operations on them are called \emph{succinct structures} \cite{RamanRR2002,Jacobson1989,Patrascu2008,Mitzenmacher2002}. Thus, we provide a succinct structure for feedforward neural networks, which may fit on-device and may enable scaling of DLaaS in the cloud.

Over the past couple of years, there has been growing interest in compact representations of neural networks \cite{GongLYB2014_arXiv, CourbariauxBD2015, GuptaAGN2015, ChenWTWC2015, LuSS2016, HanMD2016, KimPYCYS2016, LinTA2016, LaneBGFJQK2016, HintonVD2015_arXiv}, largely focused on lossy representations, see \cite{ChengWZZ2018} for a recent survey of developed techniques including pruning, pooling, and factoring.  These works largely lack strong information-theoretic foundations and may discretize real-valued weights through simple uniform quantization, perhaps followed by independent entropy coding applied to each.  It is worth noting that binary-valued neural networks (having only a network structure \cite{ChklovskiiMS2004} rather than trained synaptic weights) can often achieve high-fidelity inference \cite{CourbariauxHSEB2016_arXiv,AndriCRB2016} and that there is a view in neuroscience that biological synapses may be discrete-valued \cite{VarshneySC2006}.  

Neural networks are composed of nodes connected by directed edges.  Feedforward networks (multilayer perceptrons) have connections in one direction, arranged in layers.  An edge from node $i$ to node $j$ propagates an activation value $a_i$ from $i$ to $j$, and each edge has a synaptic weight $w_{ij}$ that determines the sign/strength of the connection.  Each node $j$ computes an activation function $g(\cdot)$ applied to the weighted sum of its inputs, which we can note is a permutation-invariant function:
\begin{equation}
a_j = g\left( \sum_{i} w_{ij}a_i\right) = g\left( \sum_{i} w_{\pi(i)j}a_{\pi(i)}\right)\mbox{,}
\label{eqn:ffnn_actv}
\end{equation}
for any permutation $\pi$. Nodes in the second layer are indistinguishable.  

Taking advantage of this permutation invariance in the structure of neural networks (previously unrecognized, e.g.~\cite{KhadiviTR2016_arXiv}) for lossless entropy coding can lead to rate reductions on top of any lossy representation technique that has been developed \cite{ChengWZZ2018}.  In particular, the structure of feedforward deep networks in layers past the input layer are unlabeled bipartite graphs where node labeling is irrelevant, much like for nonsequential data \cite{VarshneyG2006,Reznik2011b,Steinruecken2015}.  By dissipating the uncertainty in this invariance, lossless coding can compress more than universal graph compression for labeled graphs \cite{ChoiS2012}, essentially a gain of $N \log N$ bits for networks with $N$ nodes.

The first main contribution of this paper is determining the entropy limits, once the appropriate invariances are recognized. Next, to design an appropriate ``sorting'' of synaptic weights to put them into a canonical order where irrelevant uncertainty due to invariance is removed; a form of arithmetic coding is then used to represent the weights \cite{Rissanen1976, RissanenL1979}. Note that the coding algorithm essentially achieves the entropy bound. The third main contribution is an efficient inference algorithm that uses the compressed form of the feedforward neural network to calculate its output without completely decoding it, taking only $O(N)$ additional dynamic space for a network with $N$ nodes in the layer with maximum number of nodes. Finally, the paper provides experimental results of our compression and inference algorithms on feedforward neural networks trained to perform classification tasks on standard MNIST, IMDB, and Reuters datasets. 

A preliminary version of this work only dealt with universal compression and not succinctness \cite{BasuV2017}.
\subsection{Overview}
In this subsection, we describe the flow of the paper.
In Sec.~\ref{sec:FFNNS}, we discuss the basic structure and invariant properties of a feedforward neural network (multilayer perceptron), and how it can be decomposed into substructures that we call partially labeled bipartite graphs and unlabeled bipartite graphs. In Sec.~\ref{sec:RPLBG} and Sec.~\ref{sec:UBG}, we provide entropy bounds, universal compression algorithms, and inference algorithms that need not require full decompression for both partially labeled bipartite graphs and unlabeled bipartite graphs as defined in Sec.~\ref{sec:FFNNS}, respectively. Sec.~\ref{sec:DNN} provides two different compression algorithms based on the compression algorithms provided in Sec.~\ref{sec:RPLBG} and Sec.~\ref{sec:UBG} respectively. Sec.~\ref{sec:DNN} also provides an efficient inference algorithm based on the inference algorithm provided in Sec.~\ref{sec:RPLBG} that makes use of the compressed feedforward neural network for inference without fully decompressing it. Sec.~\ref{sec:EXP} provides experimental results for the compression algorithms and Sec.~\ref{sec:CONC} concludes the paper.
\section{Feedforward Neural Network Structure} \label{sec:FFNNS}
Consider a $K$-layer feedforward neural network with each (for notational convenience) layer having $N$ nodes, such that nodes in the first layer are labeled and all nodes in each of the remaining $(K-1)$ layers are indistinguishable from each other (when edges are ignored) due to the inferential invariance discussed in \eqref{eqn:ffnn_actv}. Suppose there are $m$ possible colorings of edges (corresponding to synaptic weights), and that connections from each node in a layer to any given node in the next layer takes color $i$ with probability $p_i$, $i = 0,\dots,m$, where $p_0$ is the probability of no edge. The goal is to universally find an efficient representation of this neural network structure. We will first consider optimal representation for two smaller substructures that form the layers of feedforward neural networks (after recognizing the invariance), and then return to the problem of optimally representing the full network.  The problem of neural network inference without the need to decode is interspersed in describing representations for the substructures and the full network (in Sec.~\ref{sec:RPLBG} and Sec.~\ref{sec:DNN}, we consider the problem of inference without the need to decode for partially labeled bipartite graphs and feedforward neural networks respectively).

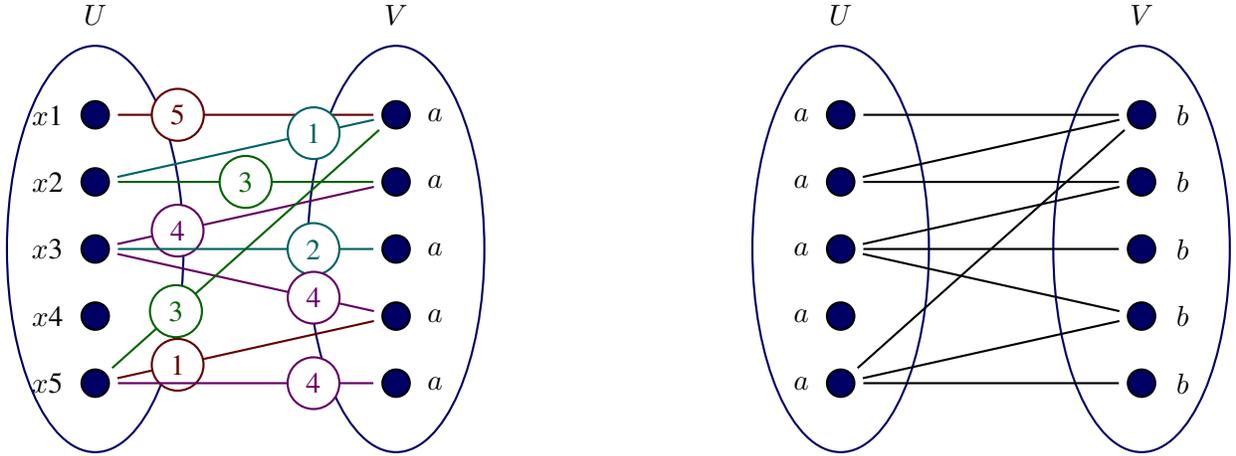
\begin{figure}
  \centering

\begin{minipage}[b]{0.4\textwidth}
 \begin{tikzpicture}[thick,
  every node/.style={draw,circle},
  fsnode/.style={fill=myblue},
  ssnode/.style={fill=mygreen},
  every fit/.style={ellipse,draw,inner sep=-2pt,text width=1.8cm},shorten >= 3pt,shorten <= 3pt
]


\begin{scope}[start chain=going below,node distance=5mm]
\foreach \i in {1,2,...,5}
  \node[fsnode,on chain] (f\i) [label=left: $x$\i] {};
\end{scope}

\begin{scope}[xshift=4cm,yshift=-0.0cm,start chain=going below,node distance=5mm]
\foreach \i in {6,7,...,9,10}
  \node[ssnode,on chain] (s\i) [label=right: $a$] {};
\end{scope}

\node [myblue,fit=(f1) (f5),label=above:$U$] {};
\node [mygreen,fit=(s6) (s10),label=above:$V$] {};

\draw (f1) -- (s6)[myred,thick] node [near start,minimum width=1mm, fill=white] {5};
\draw (s6) -- (f2)[myyellow,thick]node [near start,minimum width=1mm, fill=white] {1};
\draw (f2) -- (s7)[my2,thick]node [midway,minimum width=1mm, fill=white] {3};
\draw (s7) -- (f3)[my1,thick]node [near end,minimum width=1mm, fill=white] {4};
\draw (s8) -- (f3)[myyellow,thick]node [near start,minimum width=1mm, fill=white] {2};
\draw (f3) -- (s9)[my1,thick]node [near end,minimum width=1mm, fill=white] {4};
\draw (s9) -- (f5)[myred,thick]node [near end,minimum width=1mm, fill=white] {1};
\draw (f5) -- (s6)[my2,thick]node [near start,minimum width=1mm, fill=white] {3};
\draw (f5) -- (s10)[my1,thick]node [near end,minimum width=1mm, fill=white] {4};

\end{tikzpicture}
\end{minipage}
\hfill
\begin{minipage}[b]{0.4\textwidth}
\begin{tikzpicture}[thick,
  every node/.style={draw,circle},
  fsnode/.style={fill=myblue},
  ssnode/.style={fill=mygreen},
  every fit/.style={ellipse,draw,inner sep=-2pt,text width=1.8cm},shorten >= 3pt,shorten <= 3pt
]

\begin{scope}[start chain=going below,node distance=5mm]
\foreach \i in {1,2,...,5}
  \node[fsnode,on chain] (f\i) [label=left: $a$] {};
\end{scope}

\begin{scope}[xshift=4cm,yshift=0.0cm,start chain=going below,node distance=5mm]
\foreach \i in {6,7,...,9,10}
  \node[ssnode,on chain] (s\i) [label=right: $b$] {};
\end{scope}

\node [myblue,fit=(f1) (f5),label=above:$U$] {};
\node [mygreen,fit=(s6) (s10),label=above:$V$] {};

\draw (f1) -- (s6);
\draw (s6) -- (f2);
\draw (f2) -- (s7);
\draw (s7) -- (f3);
\draw (s8) -- (f3);
\draw (f3) -- (s9);
\draw (s9) -- (f5);
\draw (f5) -- (s6);
\draw (f5) -- (s10);

\end{tikzpicture}
\end{minipage}

\caption{(a) Partially-labeled bipartite graph with edge colors $\{0, 1, 2, 3, 4, 5\}$, where there is an edge of color $0$ between a vertex from $U$ and a vertex from $V$ if they are not connected in the figure. (b) Unlabeled bipartite graph.}
\label{fig:structures}
\end{figure}

Let us define the two aforementioned substructures: \textit{partially-labeled bipartite graphs} and \textit{unlabeled bipartite graphs}, see Fig.~\ref{fig:structures}.
\begin{definition}
A \emph{partially-labeled bipartite graph} consists of two sets of vertices, $U$ and $V$. The set $U$ contains $N$ labeled vertices, whereas the set $V$ contains $N$ unlabeled vertices. For any pair of vertices with one vertex from each set, there is a connecting edge of color $i$ with probability $p_i$, $i = 0,\dots,m$, with $p_0$ as the probability the two nodes are disconnected. Multiple edges between nodes are not allowed.
\end{definition}
\begin{definition}
An \emph{unlabeled bipartite graph} is a variation of a partially-labeled bipartite graph where both sets $U$ and $V$ consist of unlabeled vertices.
\end{definition}

In unlabeled bipartite graphs, for simplicity, in the sequel we assume there is only a single color for all nodes and that any two nodes from two different sets are connected with probability $p$.

To construct the $K$-layer neural network from the two substructures, one can think of it as made of a partially-labeled bipartite graph for the first and last layers and a cascade of $K-2$ layers of unlabeled bipartite graphs for the remaining layers. An alternative construction is also possible: the first two layers are still a partially-labeled bipartite graph but then each time the nodes of an unlabeled layer are connected, we treat it as a labeled layer, based on its connection to the previous labeled layer (i.e.\ we can label the unlabeled nodes based on the nodes of the previous layer it is connected to), and iteratively complete the $K$-layer neural network.

\section{Representing Partially-Labeled Bipartite Graphs} \label{sec:RPLBG}
We first compute the entropy bound for representing partially-labeled bipartite graphs, then introduce a universal algorithm for approaching the bound, and finally an inference algorithm that need not fully decompress to operate.

\subsection{Entropy Bound}
Consider a matrix representing the edges in a partially-labeled bipartite graph, such that each row represents an unlabeled node from $V$ and each column represents a node from $U$. A non-zero matrix element $i$ indicates there is an edge between the corresponding two nodes of color $i$, whereas a $0$ indicates they are disconnected. Observe that if the order of the rows of this matrix is permuted (preserving the order of the columns), then the corresponding bipartite graph remains the same. That is, to represent the matrix, the \emph{order of rows does not matter}. Hence the matrix can be viewed as a multiset of vectors, where each vector corresponds to a row of the matrix. Using these facts, we calculate the entropy of a partially-labeled bipartite graph. To that end, we define the following terms.

\begin{definition}
Let $\mathcal{B}(N,p)$ be a random bipartite graph model in which graphs are randomly generated on two sets of vertices, $U$ and $V$, having $N$ labeled vertices each, with edges chosen independently between any two vertices belonging to different sets with probability $p$.
\end{definition}

\begin{definition}
Let $\mathcal{B}_p(N,p)$ be a partially-labeled random bipartite graph model generating graphs in the same way as a random bipartite graph model, except that the vertices in the set $V$ in the generated graphs are unlabeled.
\end{definition}

\begin{definition}
We say that a bipartite graph, $b$ is \emph{isomorphic} to a partially labeled bipartite graph $b_p$ if $b_p$ can be obtained by removing labels from all the vertices in set $V$ of $b$, keeping all the edge connections the same. The set of all bipartite graphs, $b$, isomorphic to a partially-labeled bipartite graph, $b_p$, is represented by $I(b_p)$.
\end{definition}

\begin{definition}
The set of \emph{automorphisms} of a graph, $Aut(b)$ for $b \in \mathcal{B}(N,p)$, is defined as an adjacency-preserving permutation of the vertices of a graph; $|Aut(b)|$ denotes the number of automorphisms of a graph $b$.
\end{definition}

\begin{definition}
A graph $g$ is called \emph{asymmetric} if $|Aut(g)| = 1$; otherwise it is called \emph{symmetric}.
\end{definition}

Our proofs for entropy of random bipartite graphs follow that of \cite{ChoiS2012} for entropy of random graphs.

\begin{theorem}
\label{thm:plbge}
For large $N$, and for all $p$ satisfying $p \gg \tfrac{\ln{N}}{N}$ and $1-p \gg \tfrac{\ln{N}}{N}$, the entropy of a partially-labeled bipartite graph, with each set containing $N$ vertices and binary colored edges is 
$N^2H(p)-\log_2(N!) + o(1)$, where $H(p)=p\log_2{\tfrac{1}{p}} + (1-p)\log_2{\tfrac{1}{1-p}}$, and the notation $a \gg b$ means $b = o\left( a\right)$.
\end{theorem}
\begin{IEEEproof}
For a randomly generated bipartite graph, $b \in \mathcal{B}(N,p)$ with $k$ edges, we have
$$
P(b) = p^{k}(1-p)^{(N^2-k)}.
$$ 

Now, for each $b_p \in \mathcal{B}_p(N,p)$, there exist $|I(b_p)|$ corresponding $b \in \mathcal{B}(N,p)$ that are isomorphic to $b_p$. Hence,
$$
P(b_p) = |I(b_p)|P(b).
$$
Considering only the permutations of vertices in the set $V$, we have a total of $N!$ permutations. Given that each partially-labeled graph $b_p$ corresponds to $|I(b_p)|$ number of bipartite graphs, and each bipartite graph $b \in \mathcal{B}(N,p)$ corresponds to $|Aut(b)|$ (which is equal to $|Aut(b_p)|$) number of adjacency-preserving permutations of vertices in the graph, from \cite{HararyP1973, HararyPR1967} one gets that:
$$
N! = |Aut(b_p)|\times|I(b_p)|.
$$  

By definition, the entropy of a random bipartite graph, $H_{\mathcal{B}}$, is $N^2H(p)$ where $H(p)=p\log_2{\tfrac{1}{p}} + (1-p)\log_2{\tfrac{1}{1-p}}$. The entropy of a partially-labeled graph is:
\begin{align*}
H_{\mathcal{B}_p} &= -\sum_{b_p \in \mathcal{B}_p(N,p)} P(b_p)\log_2{P(b_p)}\\
&= -\sum_{b_p \in \mathcal{B}_p(N,p)} |I(b_p)|P(b)\log_2{(|I(b_p)|P(b))}\\
&= -\sum_{b \in \mathcal{B}(N,p)} P(b)\log_2{P(b)} -\sum_{b_p \in \mathcal{B}_p(N,p)} P(b_p)\log_2{|I(b_p)|} \\
&= -\sum_{b \in \mathcal{B}(N,p)} P(b)\log_2{P(b)} -\sum_{b_p \in \mathcal{B}_p(N,p)} P(b_p)\log_2{\tfrac{N!}{|Aut(b_p)|}} \\
&= H_{\mathcal{B}} - \log_2{N!} + \sum_{b_p \in \mathcal{B}_p(N,p)} P(b_p)\log_2{|Aut(b_p)|}\\
&= H_{\mathcal{B}} - \log_2{N!} + \sum_{b_p \in \mathcal{B}_p(N,p) \text{ is symmetric}} P(b_p)\log_2{|Aut(b_p)|} + \sum_{b_p \in \mathcal{B}_p(N,p) \text{ is asymmetric}} P(b_p)\log_2{|Aut(b_p)|} 
\end{align*}

Now \cite{KimSV2002} shows that for all $p$ satisfying the conditions in this theorem, a random graph $\mathcal{G}(N,p)$ on $N$ vertices with edges occurring between any two vertices with probability $p$ is symmetric with probability $O(N^{-w})$ for some positive constant $w$. We have stated and proved Lem.~\ref{lem:symmetry} in the Appendix to provide a similar result on symmetry of random bipartite graphs which will be used to compute its entropy.

Note that $|Aut(b_p)| = 1$ for asymmetric graphs, hence 
$$\sum_{b_p \in \mathcal{B}_p(N,p) \text{ is asymmetric}} P(b_p)\log_2{|Aut(b_p)|} = 0.$$
We know that $N! = |Aut(b_p)|\times|I(b_p)|$, hence $|Aut(b_p)| \leq N!$. Therefore, 
\begin{align*}
H_{\mathcal{B}_p}&\leq H_{\mathcal{B}} - \log_2{N!} + \sum_{b_p \in \mathcal{B}_p(N,p) \text{ is symmetric}} P(b_p)N\log_2{N}\\
&\leq H_{\mathcal{B}} - \log_2{N!} + O(\tfrac{\log_2{N}}{N^{w-1}})
\end{align*}
Hence, for any constant $w>1$,
\begin{align*}
H_{\mathcal{B}_p}&\leq N^{2}H(p) - \log_2{N!} + o(1)
\end{align*}
This completes the proof.
\end{IEEEproof}

We can also provide an alternate expression for the entropy of partially-labeled graphs with $m$ possible colors that will be amenable to comparison with the rate of a universal coding scheme.
\begin{lemma}
\label{thm:plbg}
The entropy of a partially-labeled bipartite graph, with each set containing $N$ nodes and edges colored with $m$ possibilities is 
$N^2H(p)-\log_2(N!) + E[\sum_{i=1}^{(m+1)^N}\log_2{(k_i!)}]$, where $H(p)=\sum_{i=0}^{m}p_i\log_2{\frac{1}{p_i}}$ and the $k_i$s are non-negative integers that sum to $N$.
\end{lemma}
\begin{IEEEproof}
As observed earlier, the adjacency matrix of a partially-labeled bipartite graph is nothing but a multiset of vectors. From \cite{VarshneyG2006}, we know that the empirical frequency of all elements of a multiset completely describes it. Each cell of the vector can be filled in $(m+1)$ ways corresponding to $m$ colors or no connection (color $0$), hence there can be in total $(m+1)^N$ possible vectors. The probability of a vector with the $i$th element having $K_i$ appearances is:
$$\Pr[K_i=k_i]={N\choose{k_0,k_1,\dots,k_{(m+1)^N}}}\prod_{i=1}^{(m+1)^N}\pi_i^{k_i}\mbox{.}$$
Here, $\pi_i$ is the probability of occurrence of each of the possible vectors. In the $i$th vector, let the number of edges with color $j$ be $n_j$.
Then, $\pi_i=\prod_{j=0}^{m}p_j^{n_{j}}$.
Hence, the entropy of the multiset is: 
$$
E[\log_2{\tfrac{1}{\Pr[K_i=k_i]}}] = E\left[\sum \log_2{k_i!}\right] + E\left[\sum{k_i \log_2{\tfrac{1}{\pi_i}}}\right] - \log_2 N!,
$$
and
$$
E[\sum{k_i \log_2{\tfrac{1}{\pi_i}}}]  =E\left[\sum_{(n_0,n_1,\dots,n_m)}\left(n_{(n_0,n_1,\dots,n_m)}\left(\sum_{j=0}^m n_j \log_2 \tfrac{1}{p_j}\right)\right)\right],
$$
where $n_{(n_0,n_1,\dots,n_m)}$ represents the number of vectors having $n_j$ edges of color $j$.  By linearity of expectation and rearranging terms, we get:
$$
\sum_{(n_0,n_1,\dots,n_m)}\sum_{j=0}^{m}\log_2 \tfrac{1}{p_j} E[n_j n_{(n_0,n_1,\dots,n_m)}].
$$
Now,
\begin{align*}
&\Pr[n_{(n_0,n_1,\dots,n_m)}=l]={N\choose l}{\left({{N\choose{n_0,\dots,n_m}}{\prod_{j=0}^{m}p_j^{n_j}} }\right)^l \left(1-{{N\choose{n_0,\dots,n_m}}{\prod_{j=0}^{m}p_j^{n_{j}}} }\right)^{N-l}} \\
&\quad\Rightarrow E[n_j n_{(n_0,n_1,\dots,n_m)}]=n_j N\left({{N\choose{n_0,n_1,\dots,n_m}}{\prod_{j=0}^{m}p_j^{n_{j}}} }\right)
\end{align*}
Thus,
\begin{align*}
E[\sum{k_i \log_2{\tfrac{1}{\pi_i}}}] &=\sum_{j=0}^{m}N \log_2 \tfrac{1}{p_j}\left(\sum_{(n_0,n_1,\dots,n_m)} n_j \left({{N\choose{n_0,n_1,\dots,n_m}}{\prod_{j=0}^{m}p_j^{n_{j}}} }\right)\right) \\
&=\sum_{j=0}^{m}{N^2 p_j \log_2{\tfrac{1}{p_j}}} = N^2H(p).
\end{align*}
\end{IEEEproof}

\subsection{Universal Lossless Compression Algorithm}
Next we present Alg.~\ref{alg:seq4}, a universal algorithm for compressing a partially-labeled bipartite graph based on arithmetic coding, and its performance analysis.

\begin{algorithm}[t]
\begin{algorithmic}[1]
\STATE Encode the total number of multisets in the root node of an ($m+1$)-ary tree using an integer code and initialize depth, $d=1$.
\STATE Form $m+1$ child nodes of the root node, and use arithmetic code to encode the $i$th child node with the number $x_i$, the number of vectors with $d$th cell having the $i$th color under the multinomial distribution. The vector $(x_{d,0},x_{d,1},\dots,x_{d,m})$ follows a multinomial distribution $\mathcal{M}(x_{d,0},x_{d,1},\dots,x_{d,m};N,P)$, where $P$ represents the probability vector $(p_0,p_1,\dots,p_m)$. Increase depth by 1.

\WHILE{$d \leq N$}

\FOR{each of the nodes at the current depth} 
\STATE Form $m+1$ child nodes of the current node (say, the current node is encoded with the number $\alpha$), and use arithmetic code to encode the child node of color $i$ with the number $\alpha_i$, where $\alpha_i$ represents the number of vectors with the $d$th column having color $i$ and all previous columns from $1$ to $d$ having the same colors in the same order as that of the ancestor nodes of the child node starting from the root node. Here, $(\alpha_0, \alpha_1,\dots, \alpha_m)$ follows a multinomial distribution $\mathcal{M}(\alpha_0, \alpha_1,\dots, \alpha_m;\alpha,P)$.

\ENDFOR
\STATE increase the depth by 1.
\ENDWHILE
\end{algorithmic}
\caption{Compressing a partially-labeled bipartite graph.}
\label{alg:seq4}
\end{algorithm}

\begin{lemma}
\label{thm:plbgc}
If Alg.~\ref{alg:seq4} takes $L$ bits to represent the partially-labeled bipartite graph, then $E[L] \leq  N^2H(p)-\log_2 N! +E[\sum_{i=1}^{(m+1)^N}\log_2 k_i!]+2$.
\end{lemma}
\begin{IEEEproof}
We know, for any node encoded with $\alpha$ with the encodings of its child nodes $(\alpha_0, \alpha_1,\dots, \alpha_m)$, that $(\alpha_0, \alpha_1,\dots, \alpha_m)$ is distributed as a multinomial distribution, $\mathcal{M}(\alpha_0, \alpha_1,\dots, \alpha_m;\alpha,P)$. So, using arithmetic coding to encode all the nodes, the expected number of bits required to encode all the nodes is
\begin{equation}
\label{eq:arith_lab}
E\left[\sum{\log_2{\frac{1}{\alpha ! \prod_{i=0}^{m}\frac{(p_i)^{\alpha_i}}{\alpha_i!}}}}\right].
\end{equation}
Here, the summation is over all non-zero nodes of the ($m+1$)-ary tree. Hence \eqref{eq:arith_lab} can be simplified as
$$
E[\sum \alpha_i \log_2{\tfrac{1}{p_i}}] + E[\sum \log_2{\alpha_i !}] -\log_2{N!}.
$$
When the term $E[\sum \log_2{\alpha_i !}]$ is summed over all nodes, then all terms except those corresponding to the nodes of depth $N+1$ cancel, i.e.\ $E[\sum_{i=1}^{(m+1)^N}\log_2{(k_i!)}]$.
Similarly, the term $E[\sum \alpha_i \log_2{\tfrac{1}{p_i}}]$ can be simplified as $N^2\sum_{i=0}^{m}p_i\log_2{\tfrac{1}{p_i}}$, since in the adjacency matrix of the graph, each cell can have colors from $0$ to $m$ with probability $p_i$, and for each color $i$, the expected number of cells having color $i$ is $N^2 p_i$. 
Thus, we find 
$$
E\left[\sum{\log_2{\frac{1}{\alpha ! \prod_{i=0}^{m}\frac{(p_i)^{\alpha_i}}{\alpha_i!}}}}\right]=N^2H(p)-\log_2{(N!)} +E\left[\sum_{i=1}^{(m+1)^N}\log_2{k_i!}\right].$$
Since we are using an arithmetic coder, it takes at most 2 extra bits \cite[Ch.~13.3]{CoverT2006}.
\end{IEEEproof}

\begin{theorem}
The expected compressed length generated by Alg.~\ref{alg:seq4} is within 2 bits of the entropy bound. \label{thm:plbg_comp}
\end{theorem}
\begin{IEEEproof}
The result follows from Lem.~\ref{thm:plbg} and Lem.~\ref{thm:plbgc} by comparing the entropy expression of a partially-labeled random bipartite graph with the expected length in using Alg.~\ref{alg:seq4}. 
\end{IEEEproof}

Thm.~\ref{thm:plbg_comp} states that space saving using this method can be made close to the theoretical limit. However, the theoretical limit in itself depends on the value of $N$, and hence analysis of the theoretical limit directly gives us the amount of space saving obtained. Note that the theoretical limit tells us that the space saving can be as much as $N\log{N}$ for large $N$ for partially labeled bipartite graphs with each layer having $N$ nodes, however, since the size of the graph is $O(N^2)$, the fraction of bits saved reduces as $N$ increases. On the other hand, for small values of $N$, the theoretical limit does not allow us to save around $N\log{N}$ bits. Hence there is a trade-off between the amount of bits saved and the fraction of bits saved, i.e. for small values of $N$, the fraction of bits saved is more whereas as $N$ increases, the fraction of bits saved decreases but the amount of bits saved increases.
\subsection{Inference Algorithm}
Alg.~\ref{alg:seq4} achieves near-optimal compression of partially-labeled bipartite graphs, but we also wish to use such graphs as two-layered neural networks \emph{without fully decompressing}. We next present Alg.~\ref{alg:seq5} to directly use compressed graphs for the inference operations of two-layered neural networks. Structures that take space equal to the information-theoretic minimum with only a little bit of redundancy while also supporting various relevant operations on them are called \emph{succinct structures} \cite{Patrascu2008} as defined next. 
\begin{definition}
If $L$ is the information-theoretic minimum number of bits required to store some data, then we call a structure \emph{succinct} if it represents the data in $L+o(L)$ bits, while allowing relevant operations on the compressed data.
\end{definition}
\begin{algorithm}[t]
\begin{algorithmic}[1]
\STATE \textbf{Input:} $X = [x_0,x_1,\dots,x_{N-1}]$, the input vector to the neural network, and $\mathfrak{L}$, the compressed representation of the partially-labeled bipartite graph obtained from Alg.~\ref{alg:seq4}.
\STATE \textbf{Output:} $Y = [y_0, y_1,\dots, y_{N-1}]$, the output vector of the neural network, and $\mathfrak{L}$, the compressed representation as obtained from input.
\STATE \textbf{Initialize:} $Y$ = $[y_0, y_1,\dots, y_{N-1}]$ = $[0,0,\dots,0]$, $d = 0$, the number of neurons processed at the current depth, $j = 0$, an empty queue $Q$, and an empty string $\mathfrak{L}_{1}$ which would return the compressed representation $\mathfrak{L}$ once the algorithm has executed. Let $w_i$ represent the weight corresponding to color $i$.
\STATE Enqueue $Q$ with $N$, decoded from $\mathfrak{L}$ using integer coding.
\WHILE{ $Q$ is not empty and $d\leq N-1$}
\STATE $f$ = $Q.pop()$.
\STATE $i = 0$.
\WHILE {$i \leq m$ and $f > 0$}
\STATE Using arithmetic decoding, decode the child node of $f$ from $\mathfrak{L}$ corresponding to color $i$ and store it as $c$.
\STATE Encode $c$ back in $\mathfrak{L}_1$ using arithmetic coding.
\STATE Enqueue $c$ in $Q$.   
\STATE Add $x_{d}\times w_i$ to each of $y_j$ to $y_{(j+c-1)}$.
\STATE $j = (j + c)$ mod $N$.
\IF{$j$ equals 0 and at least one non-zero node has been processed at the current depth}
\STATE $d$ = $d$ + $1$.
\ENDIF
\STATE $i = i + 1$.
\ENDWHILE
\ENDWHILE
\STATE Update the $Y$ vector using the required activation function.
\end{algorithmic}
\caption{Inference algorithm for compressed network.}
\label{alg:seq5}
\end{algorithm}

Alg.~\ref{alg:seq5} is a breadth-first search algorithm, which traverses through the compressed tree representation of the two-layered neural network and updates the output of the neural network, say $Y$, simultaneously.
Note that the $Y$ vector obtained from Alg.~\ref{alg:seq5} is a permutation of the original $\tilde{Y}$ vector obtained from the original uncompressed network. Observe that each element of $\tilde{Y}$ has a corresponding vector indicating its connection with the input to the neural network, say $X$, and when all these elements are sorted in a decreasing manner based on these connections, it gives $Y$. This happens due to the design of Alg.~\ref{alg:seq5} in giving the same $Y$ vector independent of the arrangement in $\tilde{Y}$.\footnote{Based on this invariance in the output of the compressed neural network, we can rearrange the weights of the next layers of the neural network accordingly before compressing them to get a $K$-layered neural network with the desired output as done in Sec.~\ref{sec:DNN}.}

\begin{proposition}
\label{prop:correctness}
Inference output $Y$ obtained from Alg.~\ref{alg:seq5} is a permutation of $\tilde{Y}$, the output from the uncompressed neural network representation.
\end{proposition}
\begin{IEEEproof}
We need to show that the $Y$ obtained from Alg.~\ref{alg:seq5} is a permutation of $\tilde{Y}$, obtained by direct multiplication of the weight matrix with the input vector and passed through the activation function without any compression. Say we have an $m\times 1$ vector $X$ to be multiplied with an $m\times n$ weight matrix $W$, to get the output $\tilde{Y}$, an $n\times 1$ vector. Then, $\tilde{Y} = W^{T}X$, and so the $j$th element of $\tilde{Y}$, $\tilde{Y_{j}} = \sum_{i=1}^{m}W_{j,i}^{T}x_{i}$. In Alg.~\ref{alg:seq5}, while traversing a particular depth $i$, we multiply all $Y_{j}$s with $X_{i}W_{i,j}$ and hence when we reach depth $N$, we get the $Y$ vector as required. The change in permutation of $\tilde{Y}$ with respect to $Y$ is because while compressing $W$, we do not encode the permutation of the columns, retaining the row permutation.
\end{IEEEproof}

\begin{proposition}
\label{prop:succinctness}
The additional dynamic space requirement of Alg.~\ref{alg:seq5} is $O(N)$.
\end{proposition}
\begin{IEEEproof}
It can be seen that Alg.~\ref{alg:seq5} uses some space in addition to the compressed data. The symbols decoded from $\mathfrak{L}$ are encoded into $\mathfrak{L}_{1}$, hence, the combined space taken by both of them at any point in time remains almost the same as the space taken by $\mathfrak{L}$ at the beginning of the algorithm. However, the main dynamic space requirement is because of the decoding of individual nodes, and the queue, $Q$. Clearly, the space required for $Q$, storing up to two depths of nodes in the tree, is much more than the space required for decoding a single node.

We next show that the expected space complexity corresponding to $Q$ is less than or equal to $2(m+1)N(1+2\log_{2}{(\frac{m+2}{m+1})})$ using Elias-Gamma integer codes (with a small modification to be able to encode $0$ as well) for each entry in $Q$. Note that $Q$ has nodes from at most two consecutive depths, and since only the child nodes of non-zero nodes are encoded, and the number of non-zero nodes at any depth is less than $N$, we can have a maximum of $2(m+1)N$ nodes encoded in $Q$. Let $\alpha_{0},..., \alpha_{k}$ be the values stored in the child nodes of non-zero tree nodes at some depth $d$ of the tree, where $k \leq (m+1)N$. If $k < (m+1)N$, let $\alpha_{k+1},..., \alpha_{(m+1)N}$ be all zeros. 
Let $S$ be the total space required to store $Q$. Using integer codes, we can encode any positive number $x$ in $2\log_{2}{(x)} + 1$ bits, and to allow $0$, we need $2\log_{2}{(x+1)} + 1$ bits\cite{Elias1975}.  Thus, the arithmetic-geometric inequality implies
$$
S \leq 2 \left(\sum_{i = 0}^{(m+1)N} {2\log_2{(\alpha_{i}+1)}+1}\right) \leq 2N(m+1) + 4N(m+1)\log_{2}{(\tfrac{m+2}{m+1})}.
$$
\end{IEEEproof}
\begin{theorem}
The compressed representation formed in Alg.~\ref{alg:seq4} is succinct in nature.
\label{thm:succinctness}
\end{theorem}
\begin{IEEEproof}
From Prop.~\ref{prop:correctness} and Prop.~\ref{prop:succinctness} we know that the additional dynamic space required for Alg.~\ref{alg:seq5} is $O(N)$, while the entropy of a partially-labeled bipartite graph is $O(N^{2})$.
Thus, from the definition of succinctness, it follows that the structure is succinct.
\end{IEEEproof}
Next, we will find the time complexity of Alg.~\ref{alg:seq5}.
\begin{proposition}
\label{prop:time_complexity}
The time complexity of Alg.~\ref{alg:seq5} is $O(mN^2)$.
\end{proposition}
\begin{IEEEproof}
The time taken by Alg.~\ref{alg:seq5} is the sum of time taken while decompressing the nodes and then compressing back each node of the tree, and computing the output using the decompressed node values. Assuming that multiplication takes constant time, the time taken for performing computations to get the output $Y$ is $O(N^2)$, since for any $i \in \{0,\ldots, N-1\}$, $x_i$ is multiplied to $y_j$ for all $j \in \{0,\ldots, N-1\}$ at most once. The task of compression and decompression essentially take the same time, hence we will simply show that the time taken for compression is $O(mN^2)$. Encoding a tree node formed in Alg.~\ref{alg:seq4} having value $K$ with its parent node having value $N$, where $K \in \{0,\ldots,N\}$, using arithmetic coding involves forming the cumulative distribution table of $K$ in $O(N)$ time and finding the interval corresponding to $K$ in the distribution table in $O(1)$ time. Hence, in the tree formed in Alg.~\ref{alg:seq4}, compressing a node having parent node with value $N$ takes time $O(N)$ time. Now, there can be at most $m+1$ nodes with any particular parent node. Thus, compression of the tree using arithmetic coding will take $O((m+1)T)$ time, where $T$ is the sum of all the node values in the tree. Also, note that the sum of node values in any layer can be at most $N$ and the depth of the tree can at most be $N$, hence $T \leq N^2$. Thus, the time complexity of Alg.~\ref{alg:seq5} is $O((m+1)N^2)$.  
\end{IEEEproof}

\section{Unlabeled Bipartite Graphs} \label{sec:UBG}
Next we consider an unlabeled bipartite graph for which we construct the adjacency matrix similarly as before, but now the possible entries in each cell will be binary corresponding to whether or not there is an edge. We first compute the entropy bound for representing unlabeled bipartite graphs, and then introduce a universal algorithm for approaching the bound.
\subsection{Entropy Bound}
Although the structure is slightly different from the previous case, it also has some interesting properties. The connectivity pattern is independent of the order of the row vectors and column vectors in this bipartite adjacency matrix. We say that a matrix has undergone a row permutation if the order of the rows of the matrix is changed while keeping the order of cells in each row unchanged. Similarly, we say that a matrix has undergone a column permutation if the order of the columns of the matrix is changed while keeping the order of cells in each column unchanged. We say that a matrix has undergone a \emph{valid} rearrangement is if it has undergone a sequence of row and column permutations. Note that under any valid rearrangement, the unlabeled bipartite graph remains unchanged.
Let $A$ represent the adjacency matrix of a bipartite graph and $a_{ij}$ be the cell in the matrix at row $i$ and column $j$. Say a valid rearrangement of $A$ transforms it to some matrix, $A'$, then, if a cell at row $i$ and column $j$ of $A$ has moved to row $k$ and column $l$ of the matrix $A'$ after transformation, then note that the set of cells in row $i$ of $A$ is the same as the set of cells in row $k$ of $A'$. We call this set of cells at row $i$ the row block corresponding to the cell $a_{ij}$, since this set of cells corresponding to $a_{ij}$ does not change under any valid rearrangement. Similarly, we call the set of cells at column $j$, the column block corresponding to the cell $a_{ij}$.

We will next show that the entropy of an unlabeled random bipartite graph is $N^2H(p)-2\log_2(N!) + o(1)$. To that end, we need the following definitions.

\begin{definition}
Let $\mathcal{B}_u(N,p)$ be an unlabeled random bipartite graph model generating graphs in the same way as a random bipartite graph model, except that the vertices in both the sets, $U$ and $V$, are unlabeled, but the sets $U$ and $V$ themselves remain labeled, i.e. two sets of unlabeled vertices having the same edge connections as that of a random bipartite graph.
\end{definition}

\begin{definition}
We say $b$ is \emph{isomorphic} to $b_u$ if $b_u$ can be formed by removing labels from all the vertices of $b$, keeping all the edge connections the same. The set of all bipartite graphs isomorphic to an unlabeled bipartite graph, $b_u$, is represented by $I(b_u)$.
\end{definition}

\begin{theorem}
\label{thm:ulbge}
For large $N$, and for all $p$ satisfying $p \gg \tfrac{\ln{n}}{n}$ and $1-p \gg \tfrac{\ln{n}}{n}$, the entropy of an unlabeled bipartite graph, with each set containing $N$ vertices and binary colored edges is 
$N^2H(p)-2\log_2(N!) + o(1)$, where $H(p)=p\log_2{\tfrac{1}{p}} + (1-p)\log_2{\tfrac{1}{1-p}}$, and the notation $a \gg b$ means $b = o\left( a\right)$.
\end{theorem}

\begin{IEEEproof}
From Thm.~\ref{thm:plbge}, we know that for a graph $b \in \mathcal{B}(N,p)$ with $k$ edges,
$$
P(b) = p^{k}(1-p)^{(N^2-k)}.
$$ 

For each $b_u \in \mathcal{B}_u(N,p)$, there exist $|I(b_u)|$ number of corresponding $b \in \mathcal{B}(N,p)$. Thus we have,
$$
P(b_u) = |I(b_u)|P(b).
$$
Considering the permutations of vertices in the sets $V$ and $U$ themselves, we have a total of $(N!)^{2}$ permutations. Given that each unlabeled graph $b_u$ corresponds to $|I(b_u)|$ number of bipartite graphs, and each bipartite graph $b \in \mathcal{B}(N,p)$ corresponds to $|Aut(b)|$ (which is equal to $|Aut(b_u)|$), we get the number of adjacency-preserving permutations of vertices in the graph, from \cite{HararyP1973, HararyPR1967}, as:
$$
(N!)^{2} = |Aut(b_u)|\times|I(b_u)|.
$$  
We also know that the entropy of random bipartite graph, $H_{\mathcal{B}}$, is $N^2H(p)$. The entropy of an unlabeled graph is:
\begin{align*}
H_{\mathcal{B}_u} &= -\sum_{b_u \in \mathcal{B}_u(N,p)} P(b_u)\log_2{P(b_u)}\\
&= -\sum_{b_u \in \mathcal{B}_u(N,p)} |I(b_u)|P(b)\log_2{(|I(b_u)|P(b))}\\
&= -\sum_{b \in \mathcal{B}(N,p)} P(b)\log_2{P(b)} -\sum_{b_u \in \mathcal{B}_u(N,p)} P(b_u)\log_2{|I(b_u)|} \\
&= -\sum_{b \in \mathcal{B}(N,p)} P(b)\log_2{P(b)} -\sum_{b_u \in \mathcal{B}_u(N,p)} P(b_u)\log_2{\tfrac{(N!)^{2}}{|Aut(b_u)|}} \\
&= H_{\mathcal{B}} - 2\log_2{N!} + \sum_{b_u \in \mathcal{B}_u(N,p)} P(b_u)\log_2{|Aut(b_u)|}\\
&= H_{\mathcal{B}} - 2\log_2{N!} + \sum_{b_u \in \mathcal{B}_u(N,p) \text{ is symmetric}} P(b_u)\log_2{|Aut(b_u)|} + \sum_{b_u \in \mathcal{B}_u(N,p) \text{ is asymmetric}} P(b_u)\log_2{|Aut(b_u)|} 
\end{align*}

We will next use a result, Lem.~\ref{lem:sym2} in the Appendix, on symmetry of random bipartite graphs to compute entropy.

Note that $|Aut(b_u)| = 1$ for asymmetric graphs and so: 
$$\sum_{b_u \in \mathcal{B}_u(N,p) \text{ is asymmetric}} P(b_u)\log_2{|Aut(b_u)|} = 0.$$
We know that ${(N!)}^2 = |Aut(b_u)|\times|I(b_u)|$, hence $|Aut(b_u)| \leq {(N!)}^2$. Therefore, 
\begin{align*}
H_{\mathcal{B}_u}&\leq H_{\mathcal{B}} - 2\log_2{N!} + \sum_{b_u \in \mathcal{B}_u(N,p) \text{ is symmetric}} P(b_u)2N\log_2{N}\\
&\leq H_{\mathcal{B}} - 2\log_2{N!} + O(\tfrac{\log_2{N}}{N^{w-1}}).
\end{align*}
Further, note that $H_{\mathcal{B}}= N^{2}H(p)$ where $H(p)=p\log_2{\tfrac{1}{p}} + (1-p)\log_2{\tfrac{1}{1-p}}$. Hence, for any constant $w>1$,
\begin{align*}
H_{\mathcal{B}_u}&\leq N^{2}H(p) - 2\log_2{N!} + o(1).
\end{align*}
\end{IEEEproof}
\subsection{Universal Lossless Compression Algorithm}
In this subsection, we provide a lossless compression algorithm for unlabeled bipartite graph which is optimal up to the second-order term. Alg.~\ref{alg:seq2} takes the adjacency matrix of an unlabeled bipartite graph as input and outputs two tree structures which are invariant to any valid rearrangement of the graph. Then these trees are compressed as follows: we perform a breadth first search on each of the trees and the child nodes of a node with value, say $N_x$, are first stored using $\lceil \log_2{(Nx+1)} \rceil$ bits and then the bit-stream produced after the completion of the breadth first search is compressed using an arithmetic encoder. Note that binomial distribution has been used for arithmetic coding, with $p$ as the probability of existence of an edge between any two nodes of the bipartite graph and $q =1-p$ as the probability that the two nodes are disconnected.

\begin{algorithm}[H]
\begin{algorithmic}[1]
\STATE Choose any cell containing 1 (call it 1-cell) from the adjacency matrix (or any cell containing 0 (0-cell) only if no 1-cell is available) and using valid rearrangements make this cell the top left element of the matrix. Call it the parent cell. Initially, all cells are unmarked.
\STATE Form two trees $t_1$ and $t_2$, and store $N$ in the root nodes of each of the trees. Initialize depth, $d= 1$.
\WHILE{depth of $t_1 \leq N+1$}
\STATE Divide every non-empty leaf node at the current depth of tree $t_1$ into two child nodes. The left child denotes the number of 1-cells that are unmarked in the column block containing the parent cell; similarly the right child denotes the remaining 0-cells that are unmarked.
\STATE Mark all unmarked cells in the column block containing the parent cell.
\STATE Remove an element from the leftmost node of the tree $t_2$.
\STATE Choose any cell from the newly formed leftmost child of the tree $t_1$ as the parent cell.
\STATE Divide all the leaf nodes at the current depth of the tree $t_2$ into two child nodes. The left child denotes the number of unmarked 1-cells in the row block containing the parent cell; similarly the right child denotes the remaining 0-cells that are unmarked.
\STATE Choose any cell from the newly formed leftmost child of the tree $t_2$ as the parent cell.
\STATE Mark all the unmarked cells in the row block containing the parent cell.
\STATE Remove an element from the leftmost node of the tree $t_1$.
\STATE Increase depth of $t_1$ and $t_2$ by 1.
\ENDWHILE
\end{algorithmic}
\caption{Compressing an unlabeled bipartite graph.}
\label{alg:seq2}
\end{algorithm}

It can be observed that the structure of the trees formed in Alg.~\ref{alg:seq2} is the same as in \cite{ChoiS2012} except that there are two trees in our algorithm and the first tree does not lose an element from the root node on its first division. Let us now define a tree structure which will be useful for the analysis of the performance of the algorithm.

\begin{definition}
Let $\mathcal{T}_{n,d,p}$ be a class of random binary trees such that any tree $T_{n,d,p} \in \mathcal{T}_{n,d,p}$ has depth $(n-1)$ and is generated in the following way: 1) The root node is assigned the value $n$ and placed at depth $0$. 2) If $d>0$, then starting from depth, $t = 0$ to $t = d-1$, divide each of the nodes with non-zero values at the current depth into two child nodes such that the sum of the values assigned to the child nodes is equal to that of the parent node (say $N$), and the left child node has value $N_1$ distributed as binomial distribution, $N_1 \sim Binomial(N,p)$. Else, if $d=0$, skip this step. 3) Starting from depth $t = d$ to $t = n-2$, subtract the value of the leftmost node with non-zero value and divide each of the non-zero nodes at the current depth into two child nodes in the same way as in the previous step using the updated node values after subtraction. That is, the sum of the values of the child nodes is equal to that of the updated value of the parent node, and the left child node has value assigned to it using binomial distribution. We write $\mathcal{T}_{n,d,p}$ as $\mathcal{T}_{n,d}$ when $p$ is clear from context, and we use the notations $T_{n,0}$ and $T_{n}$ interchangeably.
\end{definition}
Let $N_x$ be the number of elements in some node $x$ of either of the trees formed in Alg.~\ref{alg:seq2} (say $T$, where $T$ can be $t_1$ or $t_2$ formed in the algorithm). Then the total number of bits required for encoding the tree before using arithmetic coding is $\sum_{x \in T \text{ and } N_x \geq 1}\lceil \log_2{(N_x +1)} \rceil$. Define $L_1=\sum_{x \in T \text{ and } N_x > 1}\lceil \log_2{(N_x +1)} \rceil$ and $L_2=\sum_{x \in T \text{ and } N_x=1}\lceil \log_2{(N_x +1)} \rceil$. Let $\hat{L_1}$ and $\hat{L_2}$ be the length of bit-streams corresponding to $L_1$ and $L_2$ respectively after being compressed using arithmetic coding. So, the total expected bit length is $E\left[L_1\right] +E\left[L_2\right]$ before using arithmetic coding, and $E\left[\hat{L_1}\right] +E\left[\hat{L_2}\right] $ after using arithmetic coding. Now define 
\begin{align*}
a_{n,d}&=E\left[\sum_{x \in T_{n,d} \text{ and } N_x > 1}\lceil \log_2{(N_x +1)} \rceil\right]\mbox{, and}\\ 
b_{n,d}&=\sum_{x \in T_{n,d}}N_x-\sum_{x \in T_{n,d} \text{ and } N_x=1}N_x\mbox{.}
\end{align*}
Now we bound the compression performance of Alg.~\ref{alg:seq2}.  The proof for this bound is based on a theorem for compression of graphical structures \cite{ChoiS2012} and before stating our result and its proof, we recall two lemmas from there.
\begin{lemma}
For all integers $n\geq 0$ and $d\geq 0$, $$a_{n,d} \leq x_{n},$$
where $x_n$ satisfies $x_0=x_1=0$ and for $n \geq 2$,
$$
x_n=\lceil \log_2{(n+1)} \rceil + \sum_{k=0}^{n}{n \choose k}{p^k q^{n-k}(x_k+x_{n-k})}.
$$
\label{lemma:seq4}
\end{lemma}

\begin{lemma}
For all $n \geq 0$ and $d \geq 0$, $$b_{n,d} \geq y_n -\frac{n}{2},$$
such that $y_n$ satisfies $y_0 =0$ and for $n \geq 0$,
$$
y_{n+1}=n+ \sum_{k=0}^{n}{n \choose k}{p^k q^{n-k}(y_k+y_{n-k})}.
$$
\label{lemma:seq5}
\end{lemma}

\begin{theorem}
If an \textit{unlabeled bipartite graph} can be represented by Alg.~\ref{alg:seq2} in $L$ bits, then $E[L]\leq N^2H(p)-2N\log_2{(N)}+2(c+\Phi(\log_2{(N+1)}))(N+1)+o(N)$, where $c$ is an explicitly computable constant, and $\Phi(\log_2{(N+1)})$ is a fluctuating function with a small amplitude independent of $N$.
\end{theorem}
\begin{IEEEproof}
We need to find the expected value of the sum of all the encoding-lengths in all nodes of both trees. The expected value of length of encoding for both trees can be upper-bounded by an expression provided in \cite{ChoiS2012}. 

Let us formally prove that both encodings are upper-bounded by this expression. If $E[L_{t_1}]$ and $E[L_{t_2}]$ are the number of bits required to represent trees $t_1$ and $t_2$, respectively, then the following equations hold.
\begin{align*}
E[L_{t_1}]&=a_{N,1}+\tfrac{N(N+1)}{2}-b_{N,1}\mbox{,} \\
E[L_{t_2}]&=a_{N,0}+\tfrac{N(N-1)}{2}-b_{N,0}\mbox{.}
\end{align*}

Similarly, $E\left[\hat{L_{t_1}}\right]$ and $E\left[\hat{L_{t_2}}\right]$ are the number of bits required to represent trees $t_1$ and $t_2$ after using arithmetic coding, respectively.
Using Lem.~\ref{lemma:seq4} and Lem.~\ref{lemma:seq5}, and bounds on $x_n$ and $y_n$ from \cite{ChoiS2012} it follows that for any $d \geq 0$:
\begin{align*}
E\left[\hat{L_{t_1}}\right]\leq \tfrac{N(N+1)}{2}H(p)-N\log_2{N}+(c+\Phi(\log_2{(N+1)}))(N+1)+o(N)\mbox{,} \\
E\left[\hat{L_{t_2}}\right]\leq \tfrac{N(N-1)}{2}H(p)-N\log_2{N}+(c+\Phi(\log_2{(N+1)}))(N+1)+o(N)\mbox{.}
\end{align*}

Hence, the sum:
\[
E\left[\hat{L}_{t_1}\right] + E\left[\hat{L}_{t_2}\right] \leq N^2H(p)-2N\log_2{N}+2(c+\Phi(\log_2{(N+1)}))(N+1)+o(N).
\]
where $c$ is an explicitly computable constant and $\Phi(\log{(N+1)})$ is a fluctuating function with a small amplitude independent of $N$.
This completes the proof.
\end{IEEEproof}

It can be observed that by using Alg.~\ref{alg:seq2} for unlabeled bipartite graphs, we save roughly $N\log_2{N}$ bits when compared to compressing partially-labeled bipartite graph using Alg.~\ref{alg:seq4}.

\section{Deep Neural Networks}\label{sec:DNN}
Now we return to the $K$-layer neural network model from Sec.~\ref{sec:FFNNS}. First we extend the algorithm for unlabeled bipartite graph to compress $K$-layered unlabeled graph, and then store the permutation of the first and last layers. This gives us an efficient compression algorithm for a $K$-layered neural network, saving around $(K-2)\times N\log_2{N}$ bits compared to standard arithmetic coding of weight matrices. Alg.~\ref{alg:seq7} takes the feedforward neural network in the form of its weight matrices as input and outputs $K$ tree structures which are invariant to any valid rearrangement of the weight matrices. Then these trees are compressed similar to unlabeled bipartite graphs in Sec.~\ref{alg:seq2} as follows: we perform a breadth first search on each of the trees and the child nodes of a node with value, say $N_x$, are first stored using $\lceil \log_2{(Nx+1)} \rceil$ bits and then the bit-stream produced after the completion of the breadth first search is compressed using an arithmetic encoder. The binomial distribution has been used for arithmetic coding, with $p$ as the probability of existence of an edge between any two nodes of the bipartite graph and $q =1-p$ as the probability that the two nodes are disconnected. 
\subsection{Universal Lossless Compression Algorithm using Unlabeled Bipartite Graphs}
\begin{algorithm}[H]
\begin{algorithmic}[1]
\STATE Form root nodes of $K$ binary trees $t_1, t_2, \ldots, t_K$ corresponding to $K$ layers of the neural network, and store $N$ in the root node of all the trees, corresponding to the $N$ neural network nodes in each of the layers.
\STATE Initialize iteration number, $i = 1$, and layer number, $j = 1$. Let $\Gamma(j)$ represent the set of indices of trees corresponding to layers neighboring to the $j$th layer of the neural network.
\WHILE{depth of $i \leq N$}
\WHILE{depth of $j \leq K$}
\STATE \textbf{Selection:} Select a node of the neural network from layer $j$ that corresponds to one of the neural network nodes in the leftmost non-zero node of $t_j$ and subtract 1 from the leftmost non-zero node of $t_j$. 
\STATE \textbf{Division:} Divide every non-empty leaf node of the trees $t_k$ for $k \in \Gamma(j)$ into two child nodes based on the connections of the neural network nodes corresponding to the leaf nodes with the selected node in the previous step. The left child denotes the number of neural network nodes not connected to the selected node; similarly the right child denotes the neural network nodes connected to the selected node. 
\STATE Increment $j$ by 1.
\ENDWHILE
\STATE Increment $i$ by 1.
\ENDWHILE
\end{algorithmic}
\caption{Compressing a $K$-layer unlabeled graph.}
\label{alg:seq7}
\end{algorithm}

\begin{theorem}
Let $L$ be the number of bits required to represent a \textit{$K$-layer neural network model} using Alg.~\ref{alg:seq7}. Then $E[L] \leq (K-1)N^2H(p)+(K-2)NH(p)-(K-2)N\log{N}+K(c+\Phi(\log{(N+1)}))(N+1)+o(N)$, where $c$ is an explicitly computable constant, and $\Phi(\log{(N+1)})$ is a fluctuating function with a small amplitude independent of $N$.
\label{thm: dnn_bound_1}
\end{theorem}
\begin{IEEEproof}
The encoding of Alg.~\ref{alg:seq7} is similar to the encoding of Alg.~\ref{alg:seq2}. For all trees, the child nodes of any node with non-zero value $N_x$ are stored using $\left[ \log_2{N_x+1}\right]$ bits. Let the number of bits required to encode the $j$th layer be $L_{j}$. These bits are further compressed using an arithmetic coder, which gives us, say, $\hat{L}_{j}$ bits for the $j$th layer. Observe that the trees for the first and $K$th layer belong to $\mathcal{T}_{N,0}$ and $\mathcal{T}_{N,1}$ respectively. Hence, based on results from previous sections,
\[
E\left[\hat{L}_{1}\right] + E\left[\hat{L}_{K}\right] \leq N^2H(p)-2N\log_2{N}+2(c+\Phi(\log_2{(N+1)}))(N+1)+o(N).
\]

But the binary trees formed for the layers $2$ to $K-1$ are different. Instead of a subtraction from the leftmost non-zero node at each division after the first $d$ divisions as in a $\mathcal{T}_{n,d}$ type of tree, in these type of trees, let us call them $\mathcal{T}_{n,d}^{2}$ type of trees, subtraction takes place in every alternate division after the first $d$ divisions. We will follow the same procedure for compression of $t_2$ to $t_{K-1}$ as for $t_1$ and $t_K$, i.e. we will encode the child nodes of a node with value $N_x$ with $\left[ \log_2{N_x+1}\right]$ bits followed by an arithmetic coder. Now define,
\begin{align*}
a_{n,d}^{2}&=E\left[\sum_{x \in T_{n,d}^{2} \text{ and } N_x > 1}\lceil \log_2{(N_x +1)} \rceil\right]\mbox{, and}\\ 
b_{n,d}^{2}&=\sum_{x \in T_{n,d}^{2}}N_x-\sum_{x \in T_{n,d}^{2} \text{ and } N_x=1}N_x\mbox{.}
\end{align*}
We show that $a_{n,d}^{2} \leq x_n$ and $b_{n,d}^{2} \geq y_n -\frac{n}{2}$ for $x_n$ and $y_n$ as defined in Lem.~\ref{lemma:seq4} and Lem.~\ref{lemma:seq5}, respectively.  These are stated and proved as Lem.~\ref{lemma:seq8} and Lem.~\ref{lemma:seq9} in the Appendix.

Returning to the proof, since the trees $t_i$ for $i \in \{2,\dots,K-1 \}$, are all of the same type, we will have the same expected length of coding for each of them. Let the expected encoding length for a tree $t_i$ for $i \in \{2,\dots,K-1 \}$ before using arithmetic coding be $E\left[ L_{i}\right]$, and that after using arithmetic coding be $E\left[ \hat{L}_{i}\right]$. Then,
$$
E\left[ L_{i}\right] = N(N+1) + a_{N,1}^2 - b_{N,1}^2.
$$

Using upper bounds proved in Lem.~\ref{lemma:seq8} and Lem.~\ref{lemma:seq9}, from \cite{ChoiS2012}, we know that
$$
E\left[\hat{L}_i \right] \leq (N^2+N)H(p)-N\log_2{N}+(c+\Phi(\log_2{(N+1)}))(N+1)+o(N).
$$
where $c$ is an explicitly computable constant and $\Phi(\log{(N+1)})$ is a fluctuating function with a small amplitude independent of $N$. Further, since we need to store the permutation of the input and output layers, we need to store another $2\lceil N \log_2{N} \rceil$ bits. This completes the proof.
\end{IEEEproof}
\subsection{Universal Lossless Compression Algorithm using Partially-labeled Bipartite Graphs}
Now consider an alternative method to compress a deep neural network, using Alg.~\ref{alg:seq4} iteratively to achieve efficient compression. 
\begin{theorem}
Let $L$ be the number of bits required to represent a \textit{$K$-layer neural network model} through iterative use of Alg.~\ref{alg:seq4}. Then $E[L] \leq (k-1)(N^2H(p)-\log(N!) +E[\sum_{i=1}^{(m+1)^N}\log{(k_i!)}])+\log_2{N!}+c$, where $H(p)=\sum_{i=0}^{m}p_i\log{\tfrac{1}{p_i}}$, the $k_i$s are as defined in Lem.~\ref{thm:plbg}, and $c$ is a constant representing the amount of additional bits required by an arithmetic coder for initiating and finishing encoding.
\label{thm: dnn_bound_2}
\end{theorem}
\begin{IEEEproof}
If we focus only on the first two layers of the neural network model, then by Lem.~\ref{thm:plbg}, it can be compressed in less than $N^2H(p)-\log N! +E[\sum_{i=1}^{(m+1)^N}\log{(k_i!)}]$ number of bits. Once the first two layers are encoded, one can label the nodes of the second layer based on the relationship of its connectivity with the nodes of the first layer, and treat the second layer as a labeled layer. Also, the third layer is unlabeled and hence Alg.~\ref{alg:seq4} can be used again to compress the second and third layer using less than $N^2H(p)-\log N! +E[\sum_{i=1}^{(m+1)^N}\log{(k_i!)}$ number of bits. This, can be repeated until all layers are encoded. Further, we also need to store the permutation of the outer layer of the neural network, which takes an additional $\log_2{N!}$ bits.

Hence, iteratively encoding the $K$ layers gives:
\[
E[L_K] \leq (K-1)\left(N^2H(p)-\log N! +E[\sum_{i=1}^{(m+1)^N}\log{(k_i!)}]+\log_2{N!}+c\right).
\]
where c is the additional number of bits that an arithmetic coder takes to start and finish encoding.
\end{IEEEproof}
We have developed two different compression algorithms for feedforward neural networks. The compression algorithm based on partially labeled graph appears to be inefficient compared to the one based on unlabeled bipartite graph since after removing invariances from each layer, it treats the hidden layer as a labeled layer for compressing the next hidden layer, introducing some redundancy. However, both algorithms are asymptotically optimal upto the second-order term. Further, the algorithm based on the partially labeled graph is easier to implement and also enables easy updates in the compressed structure. Hence, in the next subsection, we provide an inference algorithm that makes use of compressed representation of a feedforward neural network generated using the iterative algorithm introduced in this subsection. 
\subsection{Inference Algorithm}
Inference for a $K$-layered neural network is just an extension of Alg.~\ref{alg:seq5}. In particular, the output of Alg.~\ref{alg:seq5} becomes the input for the next layers.
However, one important point to consider in compression, so as to ensure the inference algorithm of the $K$-layered neural network still works, is to appropriately rearrange the weight matrices. Note that Alg.~\ref{alg:seq5} outputs the $Y$ in a specific pattern, i.e.\ the output $Y$ is sorted based on the connections of output nodes with the input nodes; thus for the algorithm to work, we need to sort the weight matrix corresponding to the next layer accordingly before compressing them. Also, note that the last weight matrix connecting to the output layer of the $K$-layered neural network need not be compressed since it is desirable to preserve the ordering of the output layer nodes.
\begin{theorem}
The compressed structure obtained by the iterative use of Alg.~\ref{alg:seq4} is succinct.
\end{theorem}
\begin{IEEEproof}
Since each layer is computed one at a time in inference and the extra space required during the inference task of a 2-layered neural network is stored only temporarily, the extra dynamic space requirement for a $K$-layered remains the same as for the 2-layered neural network described in Alg.~\ref{alg:seq5}. Hence, the compressed representation for the $K$-layered neural network is succinct.
\end{IEEEproof}

Next we provide the time complexity for inference using Alg.~\ref{alg:seq5} iteratively and compare it with inference on an uncompressed neural network.
\begin{proposition}
\label{prop:time_complexity_k_DNN}
The time complexity of Alg.~\ref{alg:seq5} used iteratively on a $K$-layered neural network for inference is $O(mKN^2)$. The time complexity for inference on an uncompressed neural network is $O(KN^2)$
\end{proposition}
\begin{IEEEproof}
From Prop.~\ref{prop:time_complexity}, we already know that the time complexity of Alg.~\ref{alg:seq5} is $O(mN^2)$. Clearly, iteratively using Alg.~\ref{alg:seq5} $K$ times takes $O(mKN^2)$ time. Further, each layer of an uncompressed neural network requires $O(N^2)$ computation due to matrix multiplication of a vector of size $1\times N$ with a weight matrix of size $N \times N$. Hence, $K$ such layers take $O(KN^2)$ time.
\end{IEEEproof}

\section{Experiments}\label{sec:EXP}
To validate and assess our neural network compression scheme, we trained feedforward neural networks using stochastic gradient descent on three datasets, and quantized them using different quantization schemes before using our lossless compression scheme. The three datasets used are the MNIST dataset \cite{LeCunCB}, IMDB movie reviews sentiment classification dataset \cite{MaasDPHNP2011}, and the Reuters-21578 dataset \cite{Lewis1997}. The weights of each of the trained networks were uniformly quantized using 17, 33, and 65 quantization levels in the interval $[-0.16, 0.16]$.
We trained a feedforward neural network of dimension $784\times 50\times 50\times 50\times 50\times 10$ on the MNIST dataset using gradient descent to get an accuracy of $95.9\%$ on the test data. The test accuracy of the quantized networks are $87.1\%$, $94.3\%$, and $94.9\%$ for quantization levels of $17$, $33$, and $65$ respectively. 
Similarly, for the IMDB dataset, a feedforward neural network of dimension $1000 \times 128 \times 64 \times 2$ was trained which gives a test accuracy of $85.9\%$. The quantized networks give test accuracy of $77.9\%$, $84.7\%$, and $85.5\%$ for quantization levels of 17, 33, and 65 respectively.
For the Reuters-21578 dataset, we trained a feedforward neural network of dimension $1000 \times 200 \times 100 \times 46$ to get a test accuracy of $77.0\%$. The quantized networks give test accuracy of $72.9\%$, $75.9\%$, and $76.4\%$ for quantization levels of 17, 33, and 65 respectively.

The weight matrices from the second to the last layer were rearranged based on the weight matrices corresponding to the previous layers as needed for Alg.~\ref{alg:seq5} to work. These matrices, except the last matrix connected to the output, were compressed using Alg.~\ref{alg:seq4} to get the compressed network, and arithmetic coding was implemented by modification of an existing implementation.\footnote{Nayuki, ``Reference arithmetic coding,'' https://github.com/nayuki/Reference-arithmetic-coding, Nov. 2017. Our implementations can be found at \url{https://github.com/basusourya/DNN}} The compressed network performed exactly as the original quantized network (as it should have) since our compression is lossless. We observe that the extra memory required for inference is negligible when compared to the size of the compressed network. Detailed results from the experiments and dynamic space requirements are described in Tab.~\ref{tab:mnist}, Tab.~\ref{tab:imdb}, and Tab.~\ref{tab:reuters} for the MNIST, IMDB, and Reuters datasets respectively, where $H(p)$ is the empirical entropy calculated from the weight matrices. 

In these tables, the term $MNH(p)- N\log_2{N}$ represents an approximation to the theoretical bounds in Thm.~\ref{thm: dnn_bound_1} and \ref{thm: dnn_bound_2} since computing the exact bounds is difficult. The parameters ``Avg. queue length" and ``Max. queue length" represent the average and maximum dynamic space requirements for Alg.~\ref{alg:seq5} respectively. The fact that these two parameters have small values compared to the size of the network implies that inference without full decompression of the network takes marginal additional dynamic space.

Tab.~\ref{tab: uncomp vs comp} and \ref{tab: percentage time} measure the time needed for Alg.~\ref{alg:seq5}. Tab.~\ref{tab: uncomp vs comp} gives a comparison between time taken for inference using compressed and uncompressed neural networks. 
The experiments were run using a naive Python implementation on a system with 12GB RAM, Intel(R) Xeon(R) CPU @ 2.20GHz processor. Note that in Tab.~\ref{tab: uncomp vs comp} and \ref{tab: percentage time}, the neural networks are named after the data they were trained on and their quantization levels for conciseness, and that the number of parameters is the number of weights in a network. Tab.~\ref{tab: percentage time} provides the distribution of time taken by different components of Alg.~\ref{alg:seq5}. In particular, in Tab.~\ref{tab: percentage time} `$\%$ pmf computation' and `$\%$ arithmetic decoding + re-encoding' denote the percentage of time taken for computation of the pmf for arithmetic coder, and for decoding and re-encoding respectively.
Results show that time taken for making inference using compressed networks is considerably higher than corresponding uncompressed neural networks, but seemingly not impractical on an absolute scale. We further investigate the time taken by different components of Alg.~\ref{alg:seq5} in Tab.~\ref{tab: percentage time}. It can be observed that roughly $90 \%$ of the time taken in Alg.~\ref{alg:seq5} is due to arithmetic encoding/decoding and probability matrix computation.
{Arithmetic coding is an essential component of our inference algorithm }and so computational performance is also governed by efficient implementations of arithmetic coding. Efficient high-throughput implementations of arithmetic coding/decoding have been developed for video, e.g. as part of the H.264/AVC and HEVC standards~\cite{SzeB2012,SzeB2014}. Such efficient implementations would likely improve time required for our algorithms considerably.
\begin{table}
 \caption{Experiments for the MNIST dataset for Alg.~\ref{alg:seq4} and Alg.~\ref{alg:seq5}. }
  \centering
  \begin{tabular}{p{2.7cm}p{1.5cm}p{1.7cm}p{1.7cm}p{1.7cm}p{1.7cm}}
    \toprule
    Shape of weight matrix ($M\times N$) & Quantization level & {\textbf{$MNH(p)- N\log_2{N}$}} & Observed length (bits) & Avg.~queue length (bits) & Max.~queue length (bits)\\
\hline
    \multirow{3}{*}{$M = 784, N = 50$} & 17  & 152426 & 149994 & 150 & 257 \\
    								   & 33  & 188286 & 188165 & 151 & 397 \\
    								   & 65	   & 223936 & 225998 & 155 & 778 \\ \midrule
    \multirow{3}{*}{$M = 50, N = 50$}  & 17  & 9456 & 10254 & 152 & 255\\
    								   & 33  & 11743 & 11853 & 154 & 251 \\
    								   & 65	   & 14017 & 13480 & 180 & 396 \\ \midrule
    \multirow{3}{*}{$M = 50, N = 50$}  & 17  & 9456 & 10304 & 156 & 290\\
    								   & 33  & 11743 & 11892 & 165 & 336\\
    								   & 65	   & 14017 & 13465 & 194 & 569 \\ \midrule
    \multirow{3}{*}{$M = 50, N = 50$}  & 17  & 9456 & 10383 & 153 & 245\\
    								   & 33  & 11743 & 12004 & 173 &  475\\
    								   & 65	   & 14017 & 13688 & 178 &  520\\ \bottomrule
    
  \end{tabular}
 \label{tab:mnist}
\end{table}

\begin{table}
 \caption{Experiments for the IMDB dataset Alg.~\ref{alg:seq4} and Alg.~\ref{alg:seq5}.}
  \centering
  \begin{tabular}{p{2.7cm}p{1.5cm}p{1.7cm}p{1.7cm}p{1.7cm}p{1.7cm}}
    \toprule
    Shape of weight matrix ($M\times N$) & Quantization level & {\textbf{$MNH(p)- N\log_2{N}$}} & Observed length (bits) & Avg.~queue length (bits) & Max.~queue length (bits)\\
\hline
    \multirow{3}{*}{$M = 1000, N = 128$} & 17  & 436597 & 422241 & 384 & 625 \\
    								   & 33  & 562773 & 548951 & 385 & 825 \\
    								   & 65	   & 689138 & 676129 & 389 & 1379 \\ \midrule
    \multirow{3}{*}{$M = 128, N = 64$}  & 17  & 27615 & 41878 & 193 & 331\\
    								   & 33  & 35690 & 49486 & 204 & 618 \\
    								   & 65	   & 43778 & 56822 & 226 & 910 \\ \bottomrule
    
  \end{tabular}
 \label{tab:imdb}
\end{table}

\begin{table}
 \caption{Experiments for the Reuters dataset Alg.~\ref{alg:seq4} and Alg.~\ref{alg:seq5}.}
  \centering
  \begin{tabular}{p{2.7cm}p{1.5cm}p{1.7cm}p{1.7cm}p{1.7cm}p{1.7cm}}
    \toprule
    Shape of weight matrix ($M\times N$) & Quantization level & {\textbf{$MNH(p)- N\log_2{N}$}} & Observed length (bits) & Avg.~queue length (bits) & Max.~queue length (bits)\\
\midrule
    \multirow{3}{*}{$M = 1000, N = 200$} & 17  & 731756 & 711156 & 600 & 954 \\
    								   & 33  & 927898 & 909230 & 602 & 1444 \\
    								   & 65	   & 1124189 & 1107635 & 606 & 1739 \\ 
								   \midrule
    \multirow{3}{*}{$M = 200, N = 100$}  & 17  & 72664 & 87618 & 301 & 462\\
    								   & 33  & 92278 & 106227 & 307 & 822 \\
    								   & 65	   & 111907 & 124906 & 336 & 1481 \\
								   \bottomrule
    
  \end{tabular}
 \label{tab:reuters}
\end{table}

\begin{table}
\caption{{Comparison of inference time for compressed and uncompressed neural networks.}}
  \centering
 \begin{tabular}{rrrr}
	\toprule
	{\small{Network name}} & {\small{No. of parameters}} & {\small{Uncompressed inference time}} & {\small{Compressed inference time}} \\
	\midrule
	  {	
	MNIST17} &{46700} & {0.06 sec} & {2.30 sec}\\ 
	{MNIST33} &{46700}& {0.06 sec} & {2.81 sec}\\ 
	{MNIST65} &{46700}& {0.06 sec} & {3.34 sec}\\ 
	{IMDB17} &{136192}& {0.17 sec} & {6.4 sec}\\ 
	{IMDB33} &{136192}& {0.17 sec} & {7.41 sec}\\ 
	{IMDB65} &{136192}& {0.17 sec} & {8.91 sec}\\ 
	{Reuters17} &{220000}& {0.26 sec} & {10.14 sec}\\ 
	{Reuters33} &{220000}& {0.26 sec} & {12.07 sec}\\ 
	{Reuters65} &{220000}& {0.27 sec} & {14.99 sec}\\ 
	\bottomrule
\end{tabular} 
\label{tab: uncomp vs comp}
\end{table}

\begin{table}
\caption{{Percentage time taken by different components of Alg.~\ref{alg:seq5}.}}
  \centering
 \begin{tabular}{rrrrr}
	\toprule
	{\small{Network name}} & {\small{No. of parameters}}  & {\small{\% pmf computation}} & {\small{\% arithmetic decoding + re-encoding}}\\
	\midrule
	{MNIST17} &{46700} & {12}& {82}\\ 
	{MNIST33} &{46700} & {15}& {80}\\ 
	{MNIST65} &{46700} & {19}& {76}\\ 
	{IMDB17} &{136192} & {9}& {84}\\ 
	{IMDB33} &{136192} & {11}& {83}\\ 
	{IMDB65} &{136192} & {14}& {80}\\ 
	{Reuters17} &{220000} & {10}& {83}\\ 
	{Reuters33} &{220000} & {12}& {82}\\ 
	{Reuters65} &{220000} & {16}& {79}\\ 
	\bottomrule
\end{tabular} 

\label{tab: percentage time}
\end{table}

\section{Conclusion}\label{sec:CONC}
Data and models that are stored in memory and used for computation are often no longer of conventional type such as sequential texts or images, but rather could include structural data such as artificial neural networks, connectomes, phylogenetic trees, or social networks \cite{Szpankowski2012,ChoiS2012}.  Moreover there is growing interest in using neural network models for on-device intelligence and for scaling cloud-based intelligence, but high-performing deep neural networks are too large in size.  To ameliorate this storage bottleneck, we have developed lossless compression algorithms for feedforward deep neural networks that make use of their particular structural invariances in inference and can act as a final stage for other lossy techniques \cite{ChengWZZ2018}. Given that there may be limited prior knowledge on the statistics of synaptic weight and structure, our compression schemes are universal and yet asymptotically achieve novel entropy bounds. Further, we show that the proposed compressed representations are succinct and can be used for inference without complete decompression.
These compression algorithms can also be directly used in fully connected layers of other variants of neural networks, such as convolutional neural networks or recurrent neural networks.

In future work, we plan to investigate optimal quantization of real-valued synaptic weights using ideas from functional quantization \cite{ChatterjeeV2017},
but taking into account our novel form of entropy coding.

\section*{Acknowledgment}
Discussions with Avhishek Chatterjee are appreciated.

\appendix

\begin{lemma}
\label{lem:symmetry}
For all $p$ satisfying $p \gg \tfrac{\ln{N}}{N}$ and $1-p \gg \tfrac{\ln{N}}{N}$, a random partially bipartite graph is symmetric with probability $O(N^{-w})$ for any positive constant $w$.
\end{lemma}
\begin{IEEEproof}
Define $B = (\{U, V\}, E)$, a partially-labeled bipartite graph with two sets of vertices $U$ and $V$ and set of edges $E$. Let $\pi : U\cup V \rightarrow U\cup V$ be the permutation of vertices in the sets $U$ and $V$. Further, since the vertices in $U$ are labeled, we take $\pi (u)=u$ for $u \in U$. Following the definitions of \cite{KimSV2002}, for a vertex $v \in V$, we define a defect of $v$ with respect to $\pi$ to be
$$
D_{\pi}(v) = |\Gamma (\pi (v))\Delta \pi (\Gamma (v))|
$$ 
where $\Gamma (v)$ is the set of neighbors of $v$ and $\Delta$ denotes the symmetric difference of two sets, i.e., $A \Delta B = (A-B)\cup (B-A)$ for two sets $A$ and $B$. Similarly, one can define a defect of $B$ with respect to $\pi$ to be 
$$
D_{\pi}(B) = \underset{v}{\operatorname{max}}  D_{\pi}(v)
$$
and the defect of a graph $B$ can be defined as 
$$
D(B) = \underset{\pi \neq identity}{\operatorname{min}} D_{\pi}(B).
$$

A graph $B$ is symmetric if and only if $D(B) = 0$ \cite{ChoiS2012}. We will next show that $D(B) > 0$ with high probability, for which we will define a few terms and prove some preliminary results. Let $\pi$ be a permutation of vertices in $V$ such that it fixes all but $k$ vertices. Let $Z$ be the set of vertices, $\{u|\pi (u) \neq u \}$ and 
$$
X = \sum_{u \in P} D_{\pi}(u).
$$
Observe that, by definition, $D_{\pi}(u)$ is a binomially distributed random variable and $E[D_{\pi} (u)] = 2p(1-p)N$. Thus, $E[X] = 2p(1-p)kN$. Note that $X$ depends only on the edges of the graph adjacent to the vertices in $Z$, and adding or deleting any such edge $(u,v)$, for $u \in U$ and $v \in V$, will only affect $D_{\pi}(v)$ and $D_{\pi}(\pi^{-1}(v))$ each at most by 1. Since $X$ is a sum of binomially distributed random variables, each of which is formed from mutually independent binary choices with some probability, it is a random variable formed from mutually independent probabilistic binary decisions, such that say with probability $p_i$ it takes one of the two decisions. If the choices made for $X$ can be indexed by $i$, and let $c$ be a constant such that changing any such choice $i$ would change $X$ by at most $c$, then set $\sigma^{2} = c^{2} \sum_{i}p_{i} (1-p_{i})$. In our case, $c = 2$, hence, $\sigma^{2} = 4 Nkp(1-p)$. For all positive $t < \tfrac{2\sigma}{c}$, it is shown in \cite{AlonKS1997} that 
$$
P(|X- E[X]| > t\sigma) \leq 2 e^{-\tfrac{t^{2}}{4}}.
$$
Set $\epsilon = \epsilon(N,p)$ such that $\epsilon = o(1)$ and $\epsilon^{2}Np(1-p) \gg \ln{N}$. Then, for some positive constant $\alpha$
\begin{align*}
P(|X - E[X]| > \epsilon N k p (1-p)) &\leq 2 e^{-\alpha \epsilon^{2} Nkp(1-p)} \\
\implies P(|X - E[X]| \leq \epsilon N k p (1-p)) &> 1 - 2 e^{-\alpha \epsilon^{2} Nkp(1-p)}.
\end{align*}
Thus there exists a vertex $u$ in $Z$ such that $D_{\pi}(u) \geq \tfrac{(E[X] - \epsilon N k p (1-p))}{k} = (2-\epsilon)N k p (1-p)$ with probability at least $1 - 2 e^{-\alpha \epsilon^{2} Nkp(1-p)}$. Since, $D_{\pi}(B) = \underset{v}{\operatorname{max}}  D_{\pi}(v)$, we have 
\begin{align*}
P(D_{\pi}(B) \leq (2-\epsilon)Np(1-p)) &\leq 2 e^{-\alpha \epsilon^{2} Nkp(1-p)}.
\end{align*}

Note that there are ${N \choose k} k!$ possible permutations such that $N-k$ vertices are fixed; thus, there exists a permutation $\pi$ such that $D(B) < (2-\epsilon)Np(1-p)$ with probability less than 
$$\sum_{k=2}^{N}{N \choose k} k! \times (2 e^{-\alpha \epsilon^{2} Nkp(1-p)}). $$
As \cite{ChoiS2012} shows, $\sum_{k=2}^{N}{N \choose k} k! \times (2 e^{-\alpha \epsilon^{2} Nkp(1-p)})$  is $O(N^{-w})$ for any positive constant $w$. Hence, a partially-labeled random bipartite graph can be symmetric with probability at most $O(N^{-w})$.
\end{IEEEproof}

\begin{lemma}
\label{lem:sym2}
For all $p$ satisfying $p \gg \tfrac{\ln{N}}{N}$ and $1-p \gg \tfrac{\ln{N}}{N}$, a random unlabeled bipartite graph is symmetric with probability $O(N^{-w})$ for any positive constant $w$.
\end{lemma}
\begin{IEEEproof}
Define $B = (\{U, V\}, E)$, an unlabeled bipartite graph with two sets of vertices $U$ and $V$ and set of edges $E$. Let $\pi : U\cup V \rightarrow U\cup V$ be the permutation of vertices in the sets $U$ and $V$ with constraints that $\pi(u) \in U$ if $u \in U$ and similarly $\pi(u) \in V$ if $u \in V$. Following the definitions of \cite{KimSV2002}, for a vertex $v \in U\cup V$, we define a defect of $v$ with respect to $\pi$ to be
$$
D_{\pi}(v) = |\Gamma (\pi (v))\Delta \pi (\Gamma (v))|
$$ 
where $\Gamma (v)$ is the set of neighbors of $v$ and $\Delta$ denotes the symmetric difference of two sets, i.e., $A \Delta B = (A-B)\cup (B-A)$ for two sets $A$ and $B$. Similarly, one can define a defect of $B$ with respect to $\pi$ to be 
$$
D_{\pi}(B) = \underset{v}{\operatorname{max}}  D_{\pi}(v)
$$
and the defect of a graph $B$ can be defined as 
$$
D(B) = \underset{\pi \neq identity}{\operatorname{min}} D_{\pi}(B).
$$

A graph $B$ is symmetric if and only if $D(B) = 0$ \cite{ChoiS2012}. We will next show that $D(B) > 0$ with high probability, for which we will define a few terms and prove some preliminary results. Let $\pi$ be a permutation of vertices in $U \cup V$ such that it fixes all but $k$ vertices. Let $Z$ be the set of vertices, $\{u|\pi (u) \neq u \}$ and 
$$
X = \sum_{u \in P} D_{\pi}(u)
$$
Observe that, by definition, $D_{\pi}(u)$ is a binomially distributed random variable and $E[D_{\pi} (u)] = 2p(1-p)N$. Thus, $E[X] = 2p(1-p)kN$. Note that $X$ depends only on the edges of the graph adjacent to the vertices in $Z$, and adding or deleting any such edge $(u,v)$, for $u \in U$ and $v \in V$, will only affect $D_{\pi}(u)$, $D_{\pi}(\pi^{-1}(u))$, $D_{\pi}(v)$ and $D_{\pi}(\pi^{-1}(v))$ each at most by 1. Since $X$ is a sum of binomially distributed random variables, each of which is formed from mutually independent binary choices with some probability, it is a random variable formed from mutually independent probabilistic binary decisions, such that say with probability $p_i$ it takes one of the two decisions. If the choices made for $X$ can be indexed by $i$, and let $c$ be a constant such that changing any such choice $i$ would change $X$ by at most $c$, then set $\sigma^{2} = c^{2} \sum_{i}p_{i} (1-p_{i})$. In our case, $c = 4$, hence, $\sigma^{2} = 16 Nkp(1-p)$. For all positive $t < \tfrac{2\sigma}{c}$, it is shown in \cite{AlonKS1997} that 
$$
P(|X- E[X]| > t\sigma) \leq 2 e^{-\tfrac{t^{2}}{4}}.
$$
Set $\epsilon = \epsilon(N,p)$ such that $\epsilon = o(1)$ and $\epsilon^{2}Np(1-p) \gg \ln{N}$. Then, for some positive constant $\alpha$
\begin{align*}
P(|X - E[X]| > \epsilon N k p (1-p)) &\leq 2 e^{-\alpha \epsilon^{2} Nkp(1-p)} \\
\implies P(|X - E[X]| \leq \epsilon N k p (1-p)) &> 1 - 2 e^{-\alpha \epsilon^{2} Nkp(1-p)}.
\end{align*}
Thus there exists a vertex $u$ in $Z$ such that $D_{\pi}(u) \geq \tfrac{(E[X] - \epsilon N k p (1-p))}{k} = (2-\epsilon)N k p (1-p)$ with probability at least $1 - 2 e^{-\alpha \epsilon^{2} Nkp(1-p)}$. Since, $D_{\pi}(B) = \underset{v}{\operatorname{max}}  D_{\pi}(v)$, we have 
\begin{align*}
P(D_{\pi}(B) \leq (2-\epsilon)Np(1-p)) &\leq 2 e^{-\alpha \epsilon^{2} Nkp(1-p)}.
\end{align*}

Note that there are at most $\max_{k_1,k_2} {N \choose k_1}{{N \choose k_2}  k_1!k_2!}$ possible permutations such that $k_1 + k_2 = k$ and $N-k$ vertices are fixed. Also, $\max_{k_1,k_2}{N \choose k_1}{{N \choose k_2}  k_1!k_2!} \leq N^{k}$. Thus, there exists a permutation $\pi$ such that $D(B) < (2-\epsilon)Np(1-p)$ with probability less than 
$$\sum_{k=2}^{2N} N^k \times (2 e^{-\alpha \epsilon^{2} Nkp(1-p)}).$$
As \cite{ChoiS2012} shows, $\sum_{k=2}^{N} N^k \times (2 e^{-\alpha \epsilon^{2} Nkp(1-p)})$  is $O(N^{-w})$ for any positive constant $w$. It follows that $\sum_{k=2}^{2N} N^k \times (2 e^{-\alpha \epsilon^{2} Nkp(1-p)})$  is also $O(N^{-w})$ for any positive constant $w$. Hence, an unlabeled random bipartite graph can be symmetric with probability at most $O(N^{-w})$.
\end{IEEEproof}
\begin{lemma}
For all integers $n\geq 0$ and $d\geq 0$, $$a_{n,d}^{2} \leq x_{n},$$
where $x_n$ satisfies $x_0=x_1=0$ and for $n \geq 2$,
$$
x_n=\lceil \log_2{(n+1)} \rceil + \sum_{k=0}^{n}{n \choose k}{p^k q^{n-k}(x_k+x_{n-k})}.
$$
\label{lemma:seq8}
\end{lemma}
\begin{IEEEproof}
From Alg.~\ref{alg:seq7}, observe that $a_{0,d}^{2} = a_{1,d}^{2} = a_{2,0}^{2} = 0$. For $n \geq 2$, observe the following recursion relations for $a_{n,d}^{2}$:
\begin{align*}
a_{n+1,0}^{2}&=\lceil \log_2{(n+1)} \rceil + \sum_{k=0}^{n}{n \choose k}{p^k q^{n-k}(a_{k,1}^2+a_{n-k,2k+1}^2)}\mbox{, and}\\ 
a_{n,d}^{2}&=\lceil \log_2{(n+1)} \rceil + \sum_{k=0}^{n}{n \choose k}{p^k q^{n-k}(a_{k,d-1}^2+a_{n-k,2k+d-1}^2)}\mbox{.}
\end{align*}
We will prove the lemma using induction on both $n$ and $d$. For the base cases, observe that for $n = 0$ or $1$, $a_{n,d}^2 \leq x_n$. Further, for $n = 2$ and $d = 0$, $a_{2,0}^{2} \leq x_{2}$. Now, assuming that $a_{i,j}^{2}\leq x_{i}$ for $i < n$, and for $i = n$ and $j < d$, we want to show that $a_{n,d}^{2}\leq x_{n}$. We will consider the following two cases. 

\

\noindent {\bf Case} $d=0$: From the recursion relation of $x_n$ it follows that $x_{n} = \lceil \log{(n+1)} \rceil + \sum_{k=1}^{n-1}{n \choose k}{p^k q^{n-k}(x_k+x_{n-k})} + (p^n + q^n) (\sum_{k=1}^{n-1}{n \choose k}{p^k q^{n-k}(x_{k}+x_{n-k})}) + {(p^n+q^n)}^{2} (x_{n}),$
which implies that, 
\[x_{n} (1-{(p^n+q^n)}^{2}) = \lceil \log{(n+1)} \rceil + \sum_{k=1}^{n-1}{n \choose k}{p^k q^{n-k}(x_k+x_{n-k})} + (p^n + q^n) (\sum_{k=1}^{n-1}{n \choose k}{p^k q^{n-k}(x_{k}+x_{n-k})}).
\]

Similarly, $a_{n,0}^2 \leq a_{n+1,0}^2 = \lceil \log{(n+1)} \rceil + \sum_{k=0}^{n}{n \choose k}{p^k q^{n-k}(a_{k,1}^2+a_{n-k,2k+1}^2)}$ implies that,
$a_{n,0}^2 \leq \lceil \log{(n+1)} \rceil + \sum_{k=1}^{n-1}{n \choose k}{p^k q^{n-k}(a_{k,1}^2+a_{n-k,2k+1}^2)} + (p^n + q^n) (a_{n,1}^2)$ which in turn implies that, 

$a_{n,0}^2 \leq \lceil \log{(n+1)} \rceil + \sum_{k=1}^{n-1}{n \choose k}{p^k q^{n-k}(a_{k,1}^2+a_{n-k,2k+1}^2)} + (p^n + q^n) (\sum_{k=1}^{n-1}{n \choose k}{p^k q^{n-k}(a_{k,0}^2+a_{n-k,2k}^2)}) + {(p^n+q^n)}^{2} (a_{n,0}^2),$
which yields that
\begin{align*}
a_{n,0}^2 (1 - {(p^n+q^n)}^{2}) &\leq \lceil \log{(n+1)} \rceil + \sum_{k=1}^{n-1}{n \choose k}{p^k q^{n-k}(a_{k,1}^2+a_{n-k,2k+1}^2)}\\
 &+ (p^n + q^n) (\sum_{k=1}^{n-1}{n \choose k}{p^k q^{n-k}(a_{k,0}^2+a_{n-k,2k}^2)}).
\end{align*}
Further, 

\begin{align*}
a_{n,0}^2 (1 - {(p^n+q^n)}^{2}) &\leq \lceil \log{(n+1)} \rceil + \sum_{k=1}^{n-1}{n \choose k}{p^k q^{n-k}(x_k+x_{n-k})}\\
 &+ (p^n + q^n) (\sum_{k=1}^{n-1}{n \choose k}{p^k q^{n-k}(x_{k}+x_{n-k})})
\end{align*} 

implies that 

\begin{align*}
a_{n,0}^2 (1 - {(p^n+q^n)}^{2}) &\leq \lceil \log{(n+1)} \rceil + \sum_{k=1}^{n-1}{n \choose k}{p^k q^{n-k}(x_k+x_{n-k})}\\
 &+ (p^n + q^n) (\sum_{k=1}^{n-1}{n \choose k}{p^k q^{n-k}(x_{k}+x_{n-k})})
\end{align*}
which implies that $a_{n,0}^2 \times (1 - {(p^n+q^n)}^{2}) \leq x_{n}\times (1-{(p^n+q^n)}^{2})$.

\

\noindent {\bf Case} $d>0$: $a_{n,d}^{2}=\lceil \log_2{(n+1)} \rceil + \sum_{k=0}^{n}{n \choose k}{p^k q^{n-k}(a_{k,d-1}^2+a_{n-k,2k+d-1}^2)}$
implies that $a_{n,d}^{2} \leq \lceil \log_2{(n+1)} \rceil + \sum_{k=0}^{n}{n \choose k}{p^k q^{n-k}(x_{k}+x_{n-k})}$ which yields
$a_{n,d}^{2} \leq x_n$.
\end{IEEEproof}
\begin{lemma}
\label{lemma:seq9}
For all $n \geq 0$ and $d \geq 0$, $$b_{n,d}^{2} \geq y_n -\frac{n}{2},$$
such that $y_n$ satisfies $y_0 =0$ and for $n \geq 0$,
$$
y_{n+1}=n+ \sum_{k=0}^{n}{n \choose k}{p^k q^{n-k}(y_k+y_{n-k})}.
$$

\end{lemma}
\begin{IEEEproof}
First observe that $b_{0,d} = b_{1,d} = b_{2,0} = 0$, and for $n \geq 2$, $b_{n,d}^{2}$ forms the following recursion relation.
\begin{align*}
b_{n+1,0}^{2}&=n + \sum_{k=0}^{n}{n \choose k}{p^k q^{n-k}(b_{k,1}^2+b_{n-k,2k+1}^2)}\mbox{, and}\\ 
b_{n,d}^{2}&=n + \sum_{k=0}^{n}{n \choose k}{p^k q^{n-k}(b_{k,d-1}^2+b_{n-k,2k+d-1}^2)}\mbox{.}
\end{align*}
We will use induction on both $n$ and $d$ to prove the claim. For the base cases, clearly for $n =0$ or $n = 1$, $b_{n,d}^{2} \geq y_n - \tfrac{n}{2}$. Also for $n = 2$ and $d = 0$, $b_{n,d}^2 \geq y_n - \tfrac{n}{2}$ holds. Now, assuming that $b_{i,j}^{2}\geq y_{i} - \tfrac{i}{2}$ for $i < n$, and for $i = n$ and $j < d$, we want to show that $b_{i,j}^{2}\geq y_{i} - \tfrac{i}{2}$. We will consider the following two cases.
\

{\bf Case} $d = 0$: $b_{n,0}^{2}=(n-1)+ \sum_{k=0}^{n-1}{{n-1} \choose k}{p^k q^{n-k-1}(b_{k,1}^{2}+b_{n-k-1,2k+1})}$ which implies
$b_{n,0}^{2} \geq (n-1)+ \sum_{k=0}^{n-1}{{n-1} \choose k}{p^k q^{n-k-1}(y_k-\tfrac{k}{2}+y_{n-k-1} - \tfrac{n-k-1}{2})}$ that leads to
$b_{n,0}^{2} \geq y_n - \tfrac{n-1}{2}$ and finally, $b_{n,0}^{2} \geq y_n - \tfrac{n}{2}$.
\

{\bf Case} $d > 0$: $b_{n,d}^2=n+ \sum_{k=0}^{n}{n \choose k}{p^k q^{n-k}(b_{k,d-1}+b_{n-k,2k+d-1}^2)}$ implies $b_{n,d}^2 \geq y_{n+1} - \tfrac{n}{2}$.
From \cite{ChoiS2012}, we know that $y_{n+1} \geq y_{n}$, and so $b_{n,d}^2 \geq y_{n} - \tfrac{n}{2}$.
\end{IEEEproof}

\bibliographystyle{IEEEtran}
\bibliography{abrv,conf_abrv,lrv_lib}
\end{document}